\documentclass[twocolumn]{aastex631}

\usepackage{float}
\usepackage{wrapfig}
\usepackage{amsmath}
\usepackage{graphicx}
\restylefloat{figure}
\newcommand{\poet}{\texttt{POET}}

\newcommand\Msun{$M_{\odot}$} %
\newcommand\Rsun{$R_{\odot}$} %
\newcommand\Lsun{$L_{\odot}$} %
\newcommand{\name}{TOI-1937A}
\newcommand{\pname}{TOI-1937Ab} %
\newcommand{\usp}{USP planet }

\newcommand{\ngc}{NGC 2516 }

\newcommand\ttvestimate{-0.09 sec/yr}

\newcommand\edittwo{}

\begin{document}

\title{The Ambiguous Age and Tidal History for the Ultra-Hot Jupiter TOI-1937Ab}

\author[0000-0003-4287-004X]{Alyssa R. Jankowski}
\affiliation{University of Wisconsin - Madison Department of Astronomy, 475 N. Charter St. Madison, WI 53706, USA}

\author[0000-0002-7733-4522]{Juliette Becker}
\affiliation{University of Wisconsin - Madison Department of Astronomy, 475 N. Charter St. Madison, WI 53706, USA}

\author[0000-0001-7493-7419]{Melinda Soares-Furtado}
\affiliation{University of Wisconsin - Madison Department of Astronomy, 475 N. Charter St. Madison, WI 53706, USA}

\author[0000-0001-7246-5438]{Andrew Vanderburg}
\affiliation{Department of Physics and Kavli Institute for Astrophysics and Space Research, Massachusetts Institute of Technology,
Cambridge, MA 02139, USA}
\author{Zijun He}
\affiliation{Madison West High School, 30 Ash St., Madison, WI 53726, USA}



\begin{abstract}

Ultra-short-period (USP) planets are a rare but dynamically significant subset of the exoplanet sample, and understanding their dynamical histories and migration processes is necessary to build a complete picture of the outcomes of planet formation. In this work, we present an analysis of system age constraints and the impact of tidal evolution in the TOI-1937A system, a component of a large-separation stellar binary with an ambiguous age constraint that hosts a massive ($>2 M_{Jup}$) USP planetary companion. Through a suite of tidal evolution simulations and analysis of the transit timing variations present in the photometric data, we find that the ultra-hot Jupiter TOI-1937Ab is likely undergoing orbital decay driven by tidal interactions, and we place an {observational} upper limit on its decay rate \edittwo{of $|\dot{P}| < 0.09$ sec/yr.} We consider three different hypotheses for the system age based on three distinct methods of age estimation. These three age limits are complemented by indirect evidence of the age of the star that comes from our dynamical and transit timing analyses. We discuss the possibility that future data will provide more concrete constraints on the tidal parameters of TOI-1937Ab and its host star. 

\end{abstract}



\section{Introduction} \label{sec:intro}
Ultra-short-period (USP) planets, defined as those with orbital periods of $\leq$ 1 day \citep{Winn2018}, are relatively rare in the exoplanet sample \citep[orbiting only around 0.5\% of stars;][]{SanchisOjeda2014} but provide significant constraints on the processes of planet formation. 
USP planets are located interior to the inferred truncation radius of standard T Tauri disks \citep{Martin2011}, providing a challenge to traditional pictures of planet formation and disk-driven migration \citep{Kley2012, Pollak1996}. 
Mechanisms proposed to explain the orbits of USP planets include those that invoke planet-planet interactions \citep{Pu2019}, including secular chaos \citep{Petrovich2019} and obliquity-mediated tidal dissipation \citep{Millholland2020}, or planet-disk interactions \citep{Becker2021}.
The variability of USP planet occurrence rate with stellar age can help differentiate between these mechanisms: if the occurrence rate of USP planets is increasing by stellar age, that supports mechanisms that operate over system lifetimes \citep[i.e.,][]{Petrovich2019}; if USP occurrence rates are constant with system age, that supports mechanisms that place them in their final orbits during the disk phase \citep{Becker2021} or early in the star's life \citep{Lee2017}; finally, if the occurrence rate of USP planets is decreasing by stellar age, that suggests that processes such as tides that destroy USP planets \citep{AlvaradoMontes2021} are important in shaping the demographics of this sample. 

Efforts to constrain the occurrence rates of USP planets as a function of stellar age face two main challenges: 1) detecting planets around young stars is more difficult due to increased stellar variability \citep{2014MNRAS.438.2717J}, and 2) accurate age determinations are difficult at all stellar ages \citep{2010ARA&A..48..581S}. 
Despite these challenges, recent work by \citet{Schmidt2024} used the age–velocity relation to measure typical ages of USP planet-hosting stars, finding that the host stars typically had ages in the range of 5-6\,Gyr, older than the population of analogous USP-progenitors or multi-planet systems. This can be taken as evidence that USP planets arrived more recently at their observed orbital radii. 

However, to fully capture the information on USP planet occurrence rates, it is necessary to characterize the orbits and demographics of young ($<$0.5\,Gyr) systems and those that are still actively evolving. 
For younger hosts, age determinations are difficult, with some of the most accurate age determinations of young planets from from planets orbiting stars confirmed to be part of clusters for which the age is known \citep[e.g.,][]{2020AJ....160...33R,2022AJ....163..156M,Capistrant2024,2024AJ....168...41T}

Age determination of non-cluster-member USP planet hosts are made more difficult because tidal `spin-up' can obfuscate the true age of the system, as gyrochronological measures are a key method by which we age date systems. Additionally, interpreting the meaning of USP occurrence rates is also made more difficult by the fact that some discovered planets, (like those found in \citealt{Korth2023, Barker2024, Vissapragada2022}), show decaying orbits and would likely not look the same (or exist at all) if observed several Gyr from now. 
For most systems, we can only obtain a limit on the decay rate from observational data, and only rarely an absolute value \citep{Maciejewski2016, Patra2017, Yee2020}. However, even upper limits on tidal decay rates can inform the dynamics at play in populations of planets: because of that, searches for tidal decay have recently become more prevalent as a way to gain a clearer understanding of the dynamical evolution of planetary systems \citep{Yeh2024, Alvarado2024, Adams2024}. If tidal decay is detected in an exoplanet system, that opens the possibility of constraining the tidal quality factors for the planet ($Q_p$) and its host star ($Q_*$) \citep{Davoudi2021, Harre2023}.

\pname\ is a massive ($>$2 M$_{Jup}$) \usp orbiting a \edittwo{1.072\Msun} \citep{Yee2023}, Sun-like host with a companion at a projected separation of 1030\,AU. {The stellar type and mass of the companion are not currently known, but the presence of a stellar companion at such a relatively small projected separation is suggestive that dynamical processes such as Kozai-Lidov interactions \citep{Kozai1962, Lidov1962} may have played a role in setting the observed orbit of \pname.} This planet is one of two currently known USP planets with a mass greater than $2\ M_{Jup}$ orbiting a host with a stellar companion, and the only planet of this type with the potential for precise age dating. It was first discovered and characterized in \cite{Yee2023}, who also conducted a preliminary investigation on whether the planet was a part of the 150\,Myr NGC 2516 stellar cluster. Though their investigation was inconclusive, with evidence being provided both toward and against membership, the potential membership of this planet would represent an incredibly important observational example for the time scales on which different planetary migration events occur. 
In this work, we examine the tidal dynamics of the TOI-1937A system, with an emphasis on determining the interplay between the age uncertainty in this system (driven by the planet's complicating effect on the age derived from gyrochronology) and the possibility of constraining the tidal quality factor $Q_*$ of its host star. Due to the massive, short period orbit of TOI-1937A b, we use tidal evolution code \poet\ \citep{Penev2014software} to evaluate the impact of dynamical tides raised on the star on the inferred properties of the system.   

In Section~\ref{sec:two}, we begin with a transit-timing variation analysis to constrain the orbital decay rate of TOI-1937A b, a gyrochronological age estimate, and a lithium abundance age estimate. In Section~\ref{sec:tides}, we present the results of a suite of tidal evolution simulations and discuss the impact of tidal evolution on our derived system parameters. In Section~\ref{sec:discussion}, we continue discussion of our tidal evolution suites with the context of our data derived parameters. Finally, we conclude in Section~\ref{sec:conclusion} with a summary of our results.




\begin{figure*}
    \centering
    \includegraphics[width=0.92\linewidth]{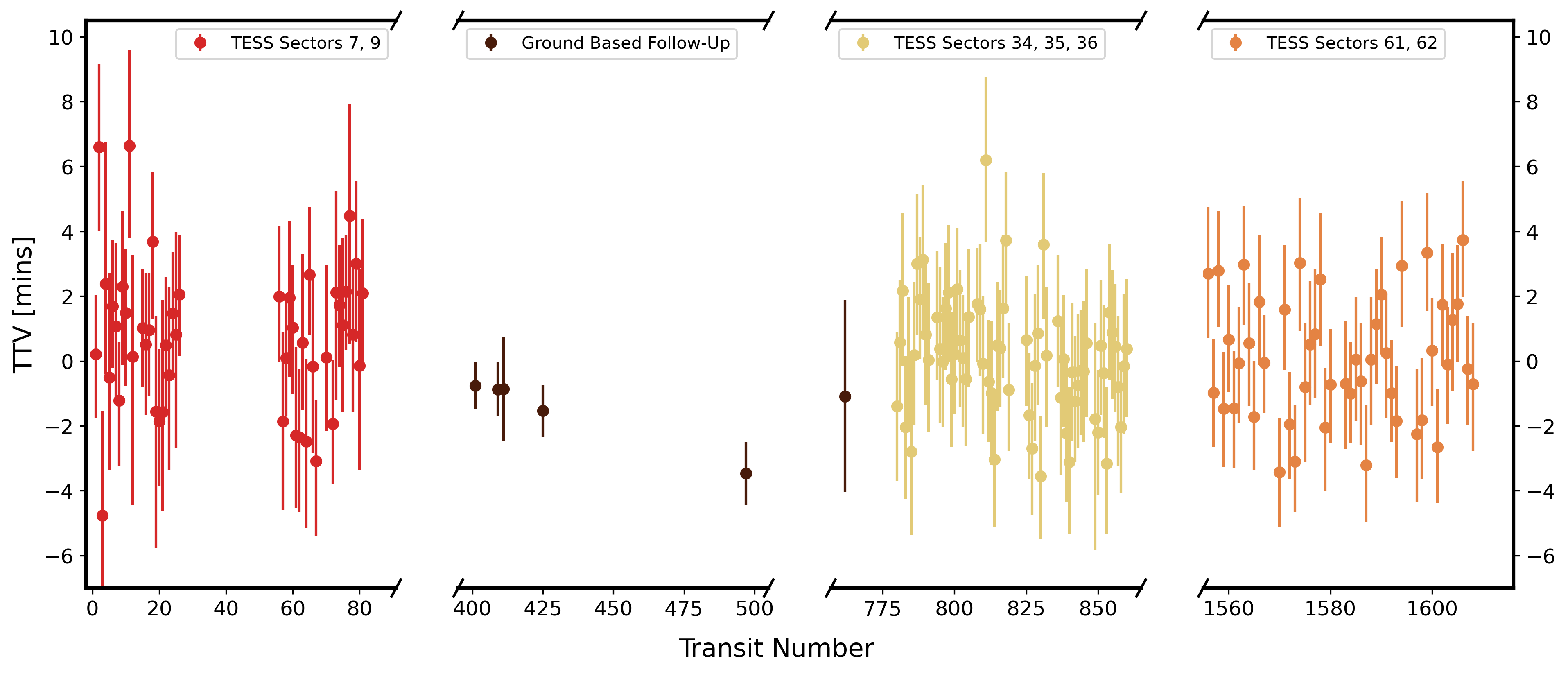}
    \caption{A plot showing transit timing variations for TOI-1937Ab measured in minutes. Points are color-coded according to the source of the data.}
    \label{fig:TTVs}
\end{figure*}

\section{Methodology and Data-Derived Parameters}
\label{sec:two}
\citet{Yee2023} provides a full EXOFASTv2 \citep{Eastman2017} fit of all available data for the TOI-1937 system and presents best-fit solutions. In Table~\ref{tab:params}, we {adopt and show} the parameters from \citet{Yee2023} relevant for our analysis. Since \citet{Yee2023} published their analysis, two additional sectors of TESS data have become available. In this section, we use these two additional sectors of TESS data, combined with the orbital solution and original data from \citet{Yee2023}, to constrain the transit timing variations (TTVs) of \pname\ and provide an updated gyrochronological estimate of the stellar rotation rate. The goal of these analysis is to provide estimates of the tidal decay rate and stellar rotation rate, which will allow us to perform a dynamical analysis on how tidal interactions are affecting the evolution of this system. 

For these analyses, we used 7 sectors of data from the Transiting Exoplanet Survey Satellite (TESS). 
The photometry for sectors 7 and 9, which were from the TESS Prime Mission, are from the full-frame images for this sector, as TOI-1937A was not a preselected star for 2 minute observations and data is available at a 30-minute cadence. For TESS' first Extended Mission (EM1), TOI-1937Ab was identified as a planet candidate, and was selected for 2-minute cadence observations in sectors 34, 35, and 36. 
We also include the ground-based follow-up transit observations published in \citet{Yee2023}. Our errors for the ground-based follow up were derived individually for each transit by using the residuals of a  {\textit{BAsic Transit Model cAlculatioN}, (\texttt{batman}, \citealt{Kreidberg2015})},  transit model with the best-fit transit parameters from \citet{Yee2023}.
In TESS Extended Mission 2 (EM2), TOI-1937Ab was selected for 20-second cadence observations and observed in sectors 61 and 62. 
All the data we use was first reported in \citet{Yee2023}, barring the data from sectors 61 and 62, which were taken after their paper was published.

Because TOI-1937A has a clear rotation signal in the light curve, we performed custom processing to account for that signal in the systematics correction. For the 30 minute cadence FFI data, we followed the procedure of \citet{Vanderburg2016ApJS} and extracted light curves from 20 different photometric apertures. We applied a systematics correction similar to \citet{Vanderburg2019ApJL}, simultaneously fitting the light curve with a basis spline (with break points every 0.5 days), time series of the moments of the spacecraft quaternion measurements within each exposure, and the background flux time series. We applied a similar systematics correction to the 2-minute and 20-second data, but started from the Simple Aperture Photometry (SAP) light curves produced by the TESS SPOC pipeline \citep{Jenkins2016SPIE}. 

\begin{deluxetable}{lccc}[htb!]
\label{tab:params}
\centering
\tabletypesize{\scriptsize}
\tablewidth{0pt}
\tablecaption{Properties of the planet and host star \name. \label{tab:prop}}
\tablehead{\colhead{Parameter} & \colhead{Value} & \colhead{Source} }
\startdata
\hline
\multicolumn{3}{c}{Planet Properties}\\
\hline
P (days)  & $0.94667944 \pm 0.00000047$ & \citet{Yee2023} \\
$T_c (BDJ_{TDB})$ & $2459085.91023 \pm 0.00012$ & \citet{Yee2023} \\
$a/R_{*}$ & $3.85^{+0.09}_{-0.10}$ & \citet{Yee2023} \\
$(R_p/R_{*})^2$ & $0.0141 \pm 0.0016$ & \citet{Yee2023} \\
$i$ (deg) & $77.0^{+0.44}_{-0.49}$ & \citet{Yee2023} \\
$R_p$ ($R_{J}$) & $1.247^{+0.59}_{-0.062}$ & \citet{Yee2023} \\
$M_p$ ($M_{J}$) & $2.01^{+0.17}_{-0.16}$ & \citet{Yee2023} \\
e & 0.0 (fixed) & \citet{Yee2023} \\
\hline
\multicolumn{3}{c}{Stellar Properties}\\
\hline
$P_{\rm{rot}}$ (days) & $6.5 \pm 0.6$ & This paper\\
T$_{\mathrm{eff}}$ (K) & $5814^{+91}_{-93}$ & \citet{Yee2023} \\
$\mathrm{M}_*$ (\Msun) & $1.072^{+0.059}_{-0.064}$ & \citet{Yee2023} \\
R$_*$ (\Rsun) & $1.080^{+0.025}_{-0.024}$ & \citet{Yee2023}\\ 
L$_*$ (\Lsun) & $1.202^{+0.088}_{-0.086}$ & \citet{Yee2023} \\ 
\enddata
\end{deluxetable}


\subsection{Transit Timing Variations}
To constrain the TTVs for this system, we performed a one parameter fit for each individual transit to find its center time of transit ($t_{\mathrm{tra}}$) using the Monte Carlo Markov Chain code \texttt{emcee} in conjunction with {\texttt{batman}} \citep{Foreman-Mackey2013}. The planet parameters (except for $t_0$) were set to the best-fit values reported in \citet{Yee2023} and reproduced in Table~\ref{tab:params}.  Each fit was run with 8 walkers for 5000 steps with a uniform prior on $t_0$ to within 0.4 days of the predicted time of transit from the \citet{Yee2023} solution. Once the $t_0$ values were found, we fit a \edittwo{quadratic} model to the values and the remaining residuals are the TTVs. Our fitted best-fit TTV values and uncertainties for each epoch for which data exists are plotted in Figure~\ref{fig:TTVs}.
Our maximum variation in transit time was 6.63 minutes, which occurred during transits that took place during TESS sectors 7 and 9. Due to the 1800 second cadence in these sectors, these transits often had only one or two data points during the transit. 

\begin{figure}
\includegraphics[width = \linewidth]{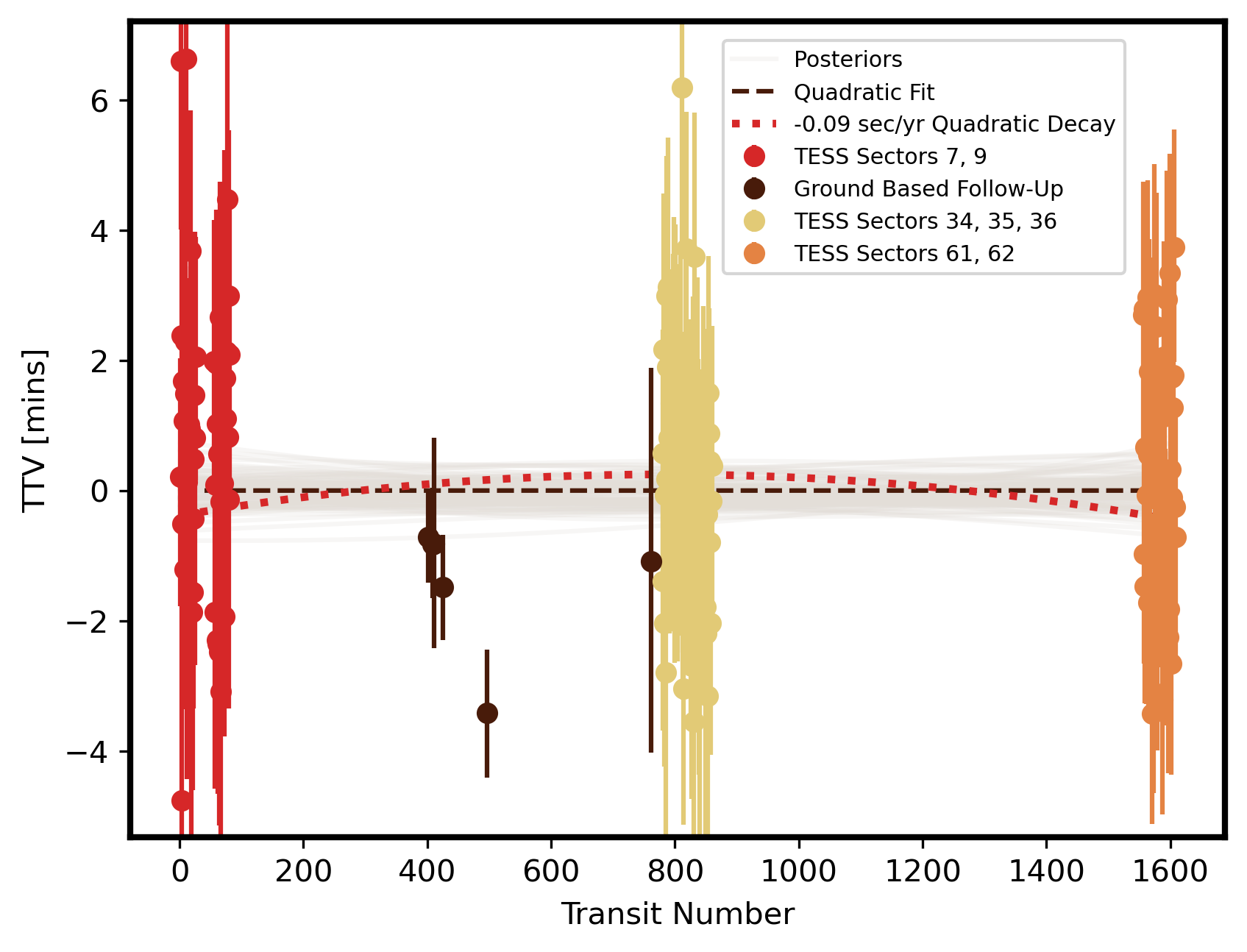}
\caption{Our derived TTVs over-plotted with 150 draws from the posterior of the quadratic fit, demonstrating the range of predicted orbital period change allowed by the data.} \label{fig:ttvpost}
\end{figure}

\begin{figure}
\includegraphics[width = \linewidth]{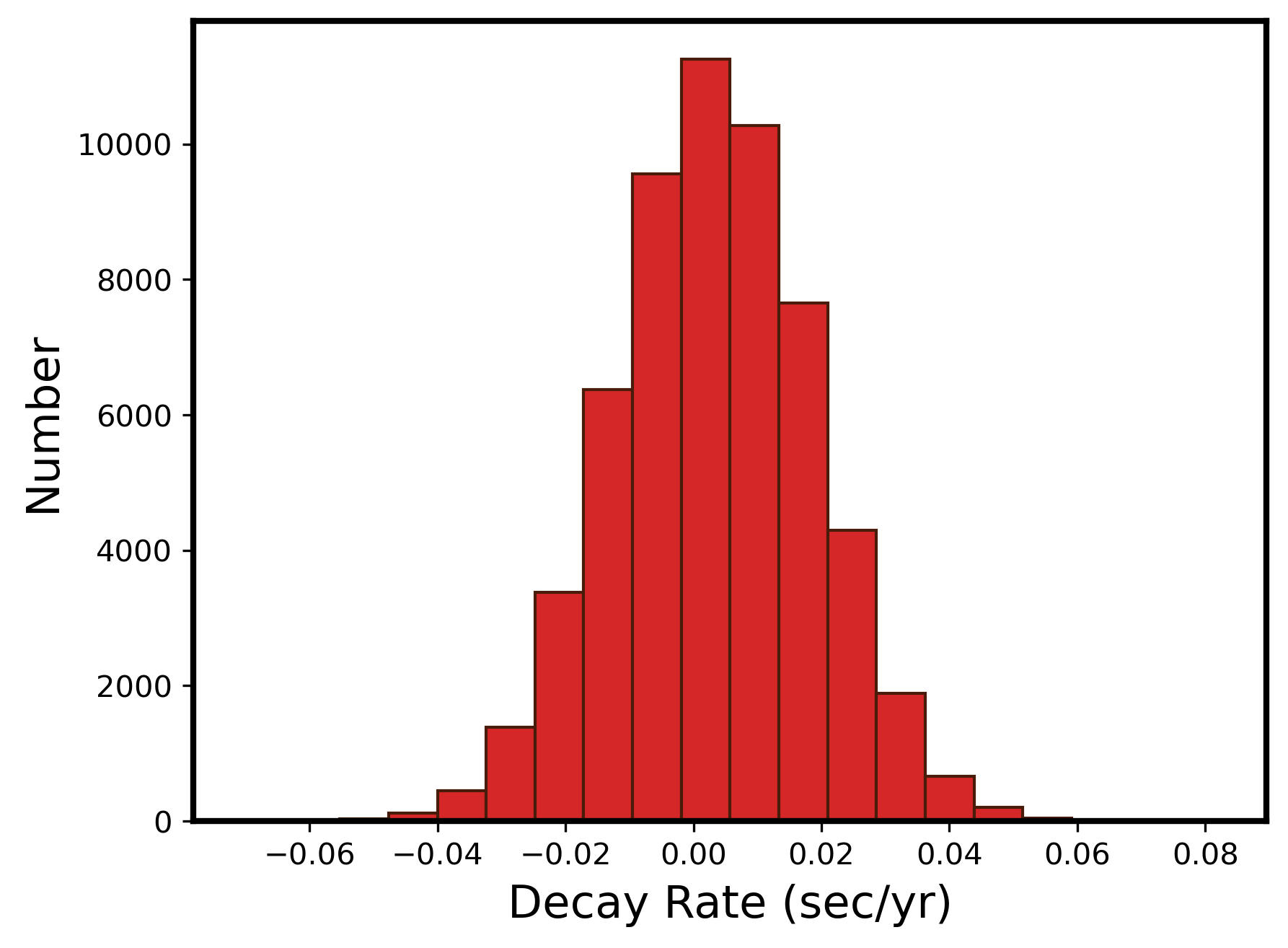}
\caption{A histogram of the posteriors for the quadratic fit to find the orbital decay in the TTV's. We note that we cannot currently exclude a slope of zero (no orbital decay) with the currently available data.}
\label{fig:ttvhist}
\end{figure}

In the full TTV curve, we found no evidence of periodic perturbations in the TTV's that would indicate the presence of another planet. 
\edittwo{To evaluate the potential for orbital decay, we fit a quadratic model using a \texttt{emcee} fit with 32 walkers and 10000 steps to our derived center times of transit \citep[as done in][]{Harre2023}:}
\begin{equation}
t_{\mathrm{tra}}(N) = t_0 + N P + \frac{1}{2} \frac{dP}{dN} N^2
\end{equation}
\edittwo{when $N$ denotes the epoch of the transit event, $t_0$ the time of transit at epoch $N =0$, $P$ the orbital period of \pname, and $dP/dN$ the orbital decay rate. The period derivative $\dot{P} = dP/dt$ is computed by $dP/dt = (1/P)\ dP/dN$. }
Using the posterior generated from this fit, \edittwo{we found a 3$\sigma$ lower limit on the change in orbital period $dP/dt$ of \ttvestimate. In Figure \ref{fig:ttvpost}, we show 150 random draws from the posterior of the quadratic model overlaid with the TTV data, along with a line corresponding to the 3$\sigma$ limit. In Figure \ref{fig:ttvhist}, we present a histogram of the posterior of the decay rate $dP/dt$.}
Further TESS observations of TOI-1937A may extend the baseline and improve this limit, but we find no current evidence of significant decay in TOI-1937Ab's orbit.

\subsection{Age Estimation}
\begin{figure}
    \includegraphics[width = \linewidth]{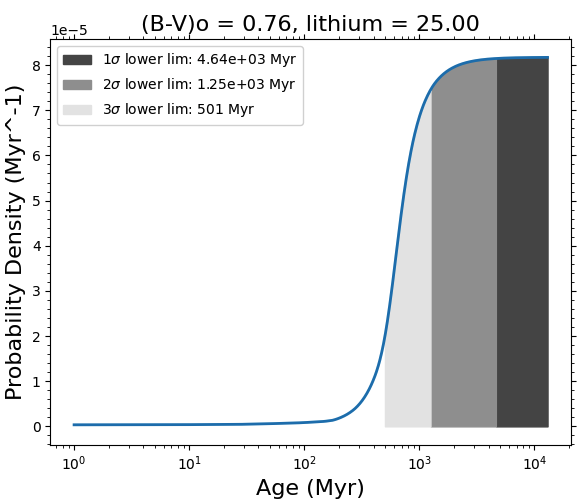}
    \caption{Probability density function for the age of \name, as calculated by the \texttt{BAFFLES} framework using the $B-V$ color and lithium EW upper limit of our target. The shaded regions illustrate the 1$\sigma$, 2$\sigma$, and 3$\sigma$ confidence intervals, corresponding to minimum ages of 4.64\,Gyr, 1.25\,Gyr, and 501\,Myr, respectively}
    \label{fig:baffles}
\end{figure}

The age of \name\ is currently debated within the literature, with some claiming that the system is a tidal tail member of the 150\,Myr \ngc open cluster, while others claim that its membership is ambiguous \citep{Kounkel2019, Yee2023}. 
However, the low upper limit implied by the Li I $6708\AA$ equivalent width (EW)  \citep[$<25\,m\AA$ at 3$\sigma$;][]{Yee2023} indicates a significantly older age than that of NGC~2516.
We investigated the age constraints imposed by this lithium EW upper limit using \texttt{BAFFLES}\footnote{\url{https://github.com/adamstanfordmoore/BAFFLES}} --- a software package capable of computing age posteriors for field stars from measurements of a target's $B-V$ color and lithium EW \citep{baffles}.
Incorporating $B-V=0.76$ \citep{Henden2016} and the star's aforementioned EW upper limit, we calculate a 3$\sigma$ minimum age of 501\,Myr (a 1$\sigma$ minimum age of 4.6\,Gyr), which is inconsistent with the measured age of the cluster. 
These findings are illustrated in Figure~\ref{fig:baffles}.

A third way to estimate the age of this system is gyrochronology. In order to estimate the photometric rotational period of TOI-1937A, we used a method similar to that laid out in \cite{Capistrant2024}. We performed a Lomb-Scargle (LS) periodogram analysis \citep{1976Ap&SS..39..447L,1982ApJ...263..835S} on the full TESS lightcurve, assisted by the Phase Dispersion Minimization (PDM) algorithm \citep{1978ApJ...224..953S} and an autocorrelation function (ACF) \citep{McQuillan2013, Shumway2005}. 

\begin{figure}
    \includegraphics[width = \linewidth]{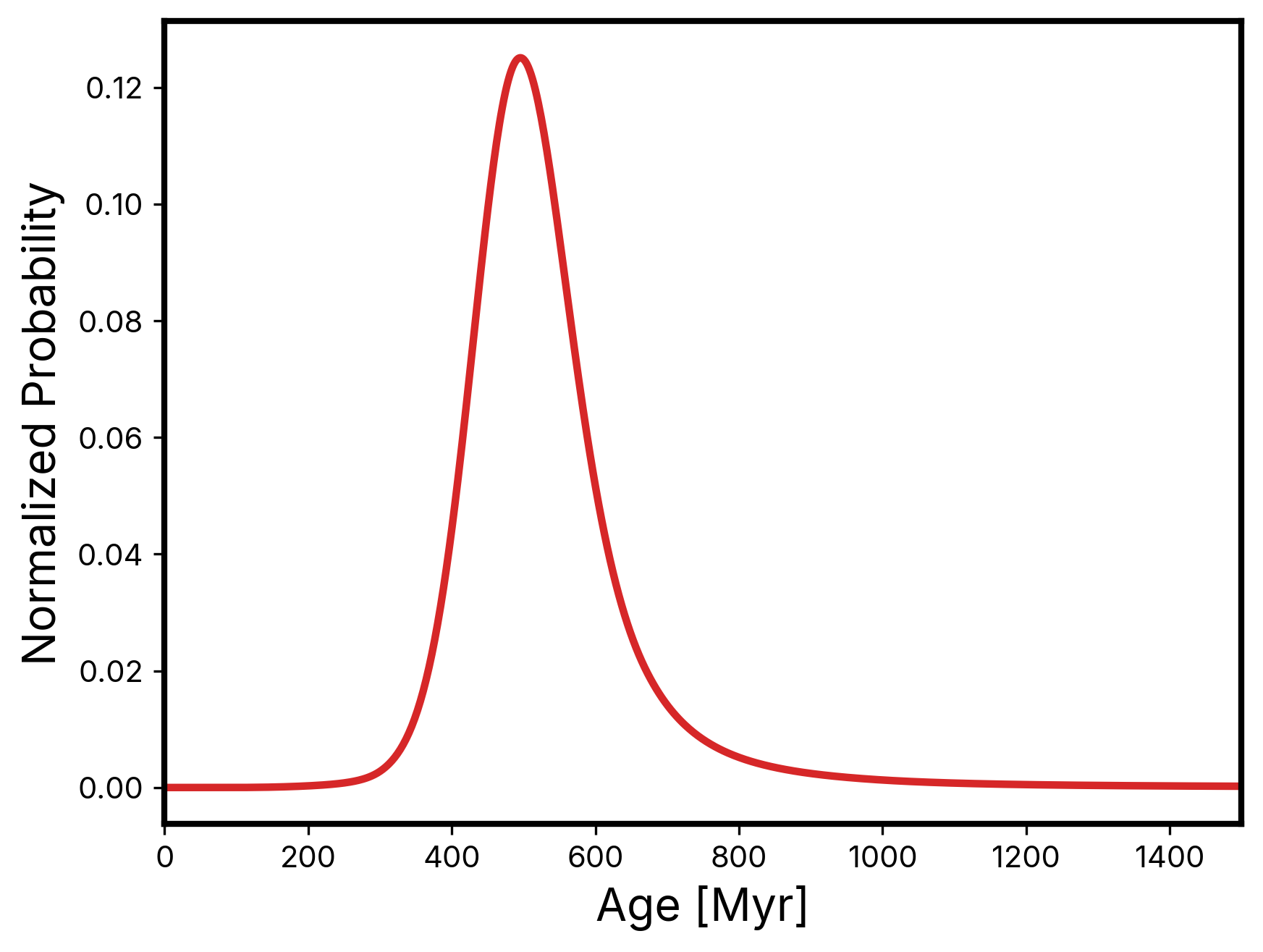}
    \caption{A plot showing the probability density function computed with \texttt{gyrointerp} \citep{Bouma2023} of the star's age based on its measured rotation period and stellar effective temperature. The best-fit age is computed to be {$510.94^{+1217.51}_{-261.51}$ Myr.}}
    \label{fig:gyro}
\end{figure}

All three period-search techniques were implemented on the full light curve, comprised of all seven sectors of TESS data along with the light curves from ground-based facilities. 
We searched for periodic signatures within a range of 0.04--30 days. We then performed a by eye search of the resulting power spectrum, autocorrelation plot, phase folded light curve, and periodogram to search for periodic structure to determine the rotation period. We determined the stellar rotation period to be 6.5 days, in agreement with the 6.6-day estimate in \cite{Yee2023}. 

After computing our estimate of the stellar rotation period, we used \texttt{gyrointerp}\footnote{\url{https://github.com/lgbouma/gyro-interp}} --- a framework that calculates the posterior probability for a star's age given its observed rotation period and effective temperature \citep{Bouma2023}.
We assumed no planet is present and used the stellar temperature found in \cite{Yee2023}.
We found an age of {$510.94^{+1217.51}_{-261.51}$\,Myr to $3\sigma$ confidence} (see Figure~\ref{fig:gyro}).

Both age estimates, whether obtained through lithium abundance methods or through gyrochronological methods, are at odds (to greater than $3\sigma$) with the 150 Myr reported age of NGC 2516. Additionally, while the $3\sigma$ minimum age estimate for the lithium abundance does allow for a 510 Myr age estimate, it is near the minimum of the 3$\sigma$ age range. All three age estimates differ wildly from each other, with the 1$\sigma$ minimum age of 4.6 Gyr and the 510 Myr  gyrochronological age estimate both indicating that TOI-1937A is a field star, and thus not belonging to NGC 2516. Though the rotation period of the star could reasonably fit within the upper edge of the spread of the rotation sequence of NGC 2516 found in \citet{Bouma2021}, the potential for tidal spin up (discussed further in Section \ref{sec:tides}) leads to gyrochronological methods being an unreliable measure of the star's age on its own.

\begin{deluxetable*}{lccc}[htb!]
\label{tab:ages}
\centering
\tabletypesize{\scriptsize}
\tablewidth{\linewidth}
\tablecaption{Age Estimates of TOI-1937A. \ref{tab:ages}}
\tablehead{\colhead{Method} & \colhead{Age Value} & \colhead{Source}}
\startdata
\multicolumn{3}{c}{} \\
Assumed Cluster Membership in NGC 2516  & 150 Myr & \citet{Bouma2021} \\
Gyrochronological Dating & $510.94^{+1217.51}_{-261.51}$\,Myr & This paper \\
Li Abundance ($1\sigma$ lower limit) & $4.64\times10^3$ Myr & This paper \\
EXOFASTv2 Median Estimate & $3.6^{+3.1}_{-2.3}\times10^3$ Myr & \citet{Yee2023} \\ \\
\enddata
\end{deluxetable*}

\section{Tidal Evolution Simulations} 
\label{sec:tides}
While gyrochronology age estimates are a common and effective tool in the literature \citep{Barnes2007, Lu2024}, in the presence of strong star-planet interactions they can produce erroneous results. 
Tidal interactions between a star and its planet can transfer angular momentum between the two, resulting in changes to the star's rotation rate over time. 
For USP planets, these interactions can be especially significant due to the strong tidal forces exerted by the planet's proximity to the host star. 
One explanation for the mismatch between the \name\ system age derived from the gyrochronology ($\sim$500 Myr), the possible cluster membership ($\sim$150 Myr), and the Li abundance ($>$4.6 Gyr at $1\sigma$ Myr) is that the planet, which has significant orbital angular momentum, could have altered stellar rotation period through its tidal evolution \citep[i.e.,][]{Ilic2024}.

A self-consistent tidal model allows for an accurate simulation of these effects by accounting for tidally-driven variations in parameters like the star's tidal dissipation factor ($Q_*$), the planet's orbital eccentricity ($e$) and semi-major axis ($a$), and the stellar rotation period $P_{rot}$. 
In this section, we use \poet\ \citep{Penev2014software, Penev2014}, a code that calculates the orbital and rotational evolution of two bodies under mutually raised tides, to model this process and assess how the rotation period of \name\ may have been affected by star-planet interactions. 


\subsection{The Numerical Model}
\poet\ provides a self-consistent framework for modeling the evolution of planetary orbits and stellar rotation, taking into account the effects of tidal interactions and stellar wind. It also incorporates the impact of core-envelope differential rotation within the stellar interior. In this section, we will outline the main equations \poet\ uses to compute the orbital evolution of the planet and the rotational evolution of its host star, which underpin our numerical explorations of the tidally-altered stellar rotation rate of \name.

First, the orbital evolution of the planet's orbit due to the (static) equilibrium tide can be written as a pair of coupled differential equations \citep{Goldreich1963,Barnes2008},
\begin{equation}
\begin{split}
\frac{da}{dt}  = \Big( &-\frac{63}{2}\frac{1}{Q' m_p}  \sqrt{GM_*^3} r_p^5 e^2 + \\
 & \frac{9}{2}\frac{\sigma}{Q_{*}'}  \sqrt{G/M_*} R_*^5 m_{p} \Big) a^{-11/2}
\end{split}
\label{eq:dadt}
\end{equation}
and 
\begin{equation}
\begin{split}
\frac{de}{dt}  = \Big( &-\frac{63}{4}\frac{1}{Q' m_p}  \sqrt{GM_*^3} r_p^5 + \\
 & \frac{171}{16}\frac{\sigma}{Q_{*}'}  \sqrt{G/M_*} R_*^5 m_{p} \Big) e\ a^{-13/2}.
\end{split}
\label{eq:dedt}
\end{equation}
Here, $Q' = 3Q / (2k_2)$ is the modified tidal dissipation factor, where $Q$ represents the tidal dissipation efficiency and $k_2$ the Love number, $m_p$ and $r_p$ denote the planetary mass and radius, respectively, $G$ is the gravitational constant, $M_*$ and $R_*$ are the stellar mass and radius, $a$ is the planetary semi-major axis, $e$ is its eccentricity, $n$ denotes the planetary mean motion, $\omega_{\text{surf}}$ the angular velocity of the stellar surface, and $\sigma = \text{sign}(2\omega_{\text{surf}} - 3n)$ attains the value 1 when the planet's orbital frequency is lower than the stellar rotation frequency and -1 in the opposite case. 

To model stellar rotation evolution over time, we must write an equation for how the stellar angular momentum changes over a star's lifetime due to two factors: (a) in response to the planet's tidally-induced orbital evolution, and (b) due to angular momentum loss caused by stellar winds \citep{Kawaler1988}. 
Following the formulation and notation of \cite{Penev2014}, we can write an expression for the change in angular momentum due to these two effects as follows:
\begin{equation}
\begin{split}
\frac{dL_{*}}{dt} &= \left( \frac{dL_{*}}{dt} \right)_{\text{tide}} + \left( \frac{dL_{*}}{dt} \right)_{\text{wind}} =  \\
&-\frac{1}{2} m_p M_* \sqrt{\frac{G}{a(M_* + m_p)}} \frac{da}{dt}  \\
&-K \omega_{\text{surf}} \min(\omega_{\text{surf}}, \omega_{\text{sat}})^2 
\left(\frac{R_*}{R_\odot}\right)^{1/2} \left(\frac{M_*}{M_\odot}\right)^{-1/2},
\end{split}
\label{eq:dLdt}
\end{equation}
where $K$ is a parameter describing the strength of the stellar magnetic wind, $\omega_{\text{sat}}$ the star's angular velocity at which the magnetic wind saturates.
Solving Equation \ref{eq:dLdt} simultaneously with Equations \ref{eq:dadt} and \ref{eq:dedt} yields the evolution over time for the planetary orbital ($a$ and $e$) and the stellar rotation rate (set by $L_*$). 

\poet\ also allows, for low-mass stars like \name, the rotation rates of the stellar core and envelope to be decoupled. \poet\ computes the orbital evolution of the planet simultaneously with the evolution of the stellar angular momenta and moments of inertia for the convective and radiative zones inside the star and the stellar mass and radius. These evolutions are based on pre-computed MESA tracks. 
Finally, \poet\ also models disk-locked stellar rotation while the protoplanetary disk is present, and the transition between non-tidally locked and tidally locked planet rotation. The details of how these calculations are implemented are given in \citet{Penev2014}. \poet\ has been used successfully in the past to assess how stellar parameters such as $Q_*$ affect the orbital evolution of planets \citep{Penev2018}, evaluate tidal dissipation in stellar binaries \citep{Penev2022}, and derive parameter constraints for $Q_p$ and $Q_*$ \citep{Mahmud2023, Patel2023}. For the work in the remainder of this section, we use the publicly available version of \poet\footnote{\url{https://github.com/kpenev/poet}}.  

\begin{figure}
\centering
   \includegraphics[width=1\linewidth]{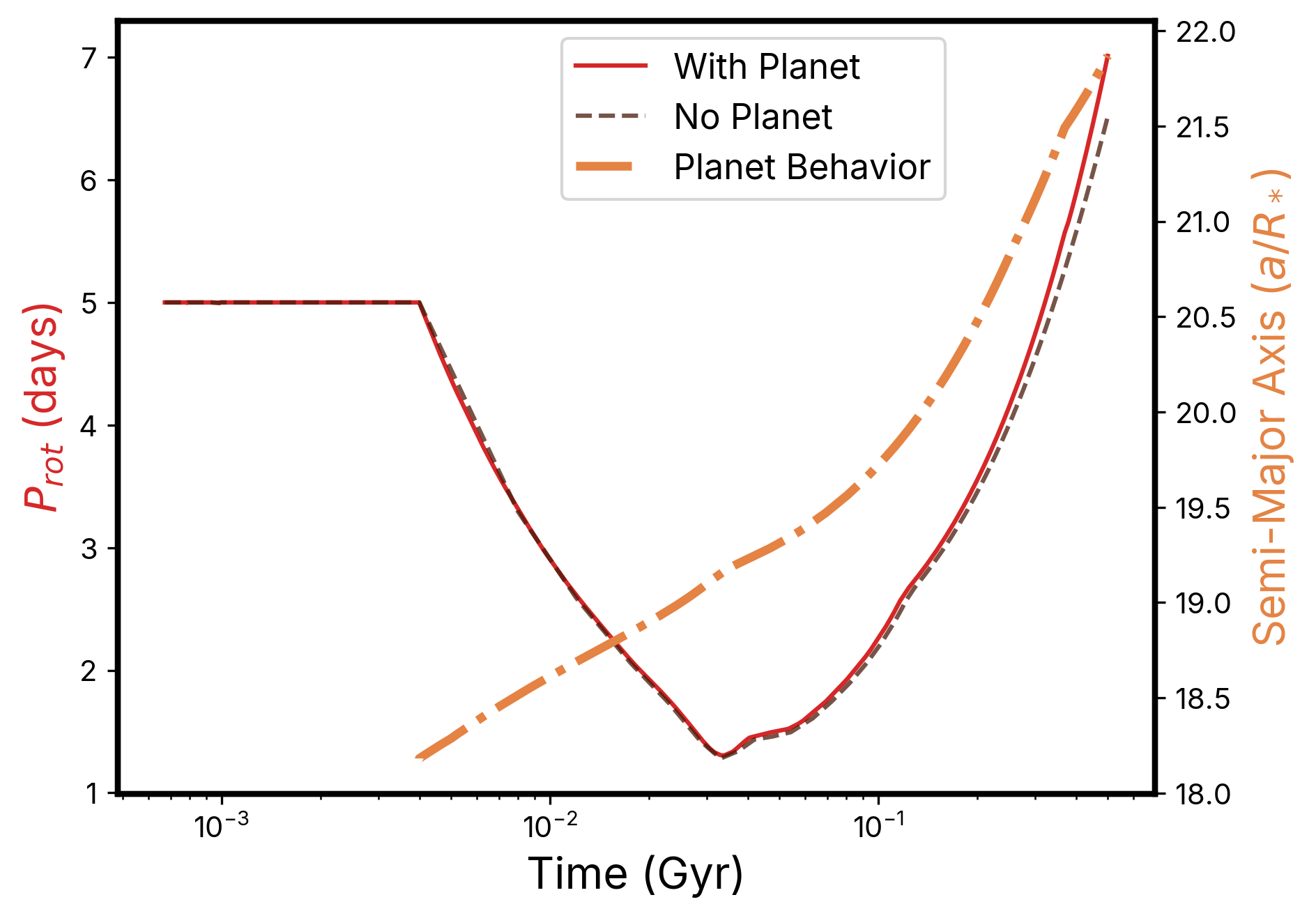}
   \includegraphics[width=1\linewidth]{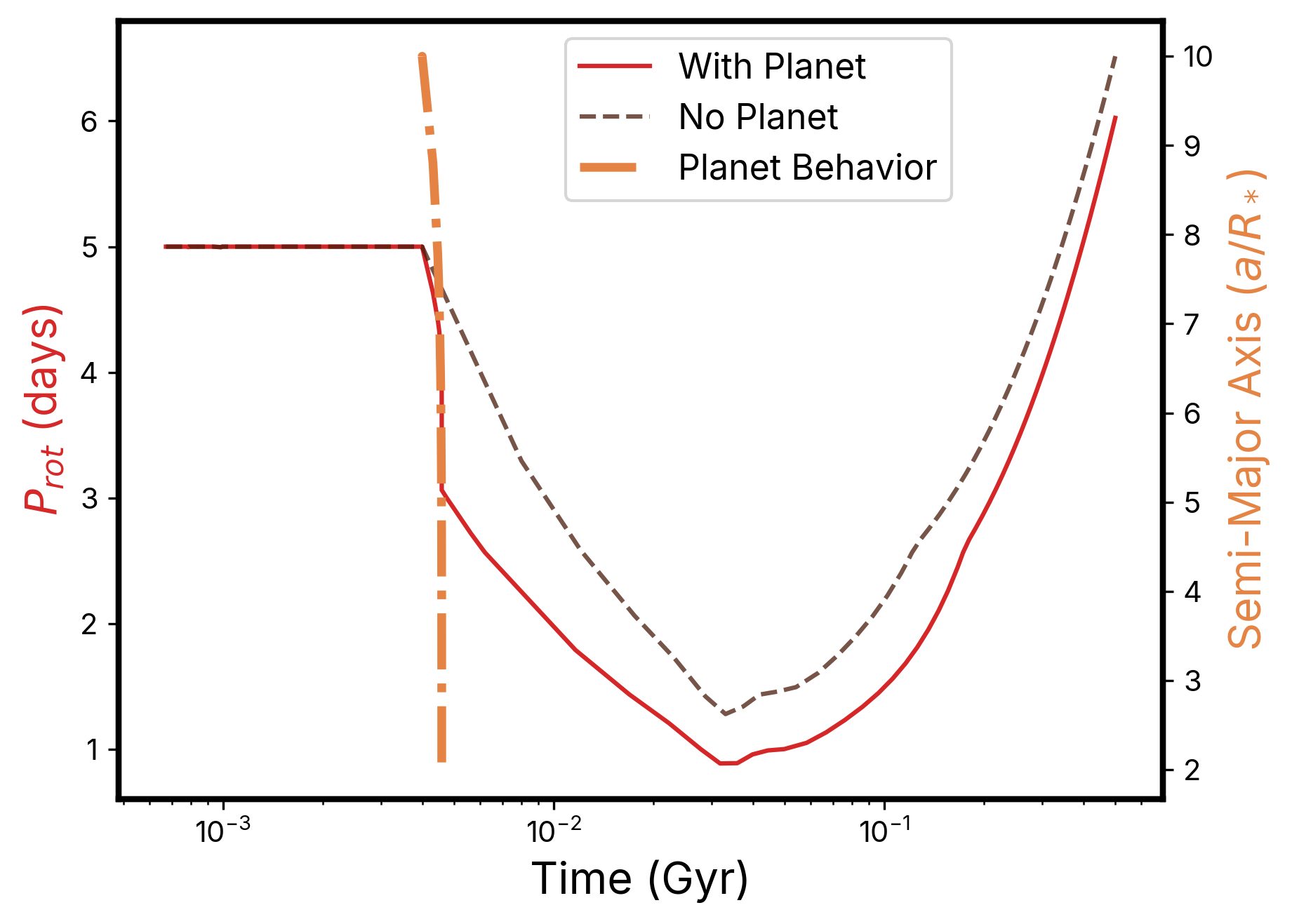}
\caption{{The two panels show two sets of planetary initial conditions that result in t}wo representative {cases of} behavior for a simulated star's rotation period {(left axis) and the semi-major axis of the orbiting hot Jupiter (right axis).} The top {panel} shows a planet with an initial orbit of $a/R_* = 18.18$ and $e = 0.18$ that survives the 5\,Gyr simulation and causes its host star's rotation period to increase compared to what would be expected with no planet present. The bottom {panel} shows a planet with an initial orbit of $a/R_* = 10.0$ and $e = 0.0$. {This second set of initial conditions results in the planet being} engulfed by its stellar host shortly into the simulation, and the host star {spins up to faster rotation speeds due to the engulfment}.}  \label{fig:Ng1} 
\end{figure}

\subsection{Stellar Rotational Evolution for \name}
In attempting to run a tidal evolution simulation that represents the true evolution of \name\ and \pname, we must choose initial orbital parameters for the planet and initial stellar parameters (such as its initial angular momentum $L_*$ and tidal quality factor $Q_{*}$) to initialize the simulation. None of these parameters are exactly known, as all we can measure from the observational data is current-day values. 
Because of this, choosing our initial simulation parameters is complicated by the fact that our age estimate of \name\ is uncertain, with different methods giving very different age estimates.
(ranging between 150 Myr and 4.5 Gyr or more). This means that the inferred $L_*$ needed to match the current-day observed rotation rate will be different for the different stellar age estimates. 

In this subsection, our goal is to determine how much the tidal migration of a planet like \pname\ could have been expected to alter the rotation rate of its host star. 
To make our exploration computationally feasible, we choose one of our computed possibilities of the stellar age to use for our simulations ($\sim500$ Myr) and choose $L_{*}$ such that the star has its observed rotation rate ($P_{rot}$ = 6.5 days) at this age.

With the stellar initial parameters chosen, we used \poet\ to compute the evolution for a set of 300 combinations of initial semi-major axis and eccentricity of the planet, spanning $10< a/R_* <100$ and $0< e <1$. For all simulations, stellar mass and radius and the planet mass were set to the best-fit values given in Table \ref{tab:params}, and the stellar quality factor set to an assumed value of $Q_{*} =10^5$. The stellar companion TOI-1937B was not modeled in the simulations, as its large projected separation of 1030 AU found in \citet{Yee2023} suggests it is dynamically decoupled. {The simulations for} \name \ begun with disk-locked rotation with a stellar rotation period of 5 days, and the disk set to dissipate at 4 Myr. At that point, the planet is assumed to form and the stellar rotation is allowed to evolve under the influence of the stellar wind and planetary tidal evolution. For each simulation, we computed the stellar rotation period at 500 Myr. 

In addition to the 300 simulations evaluating the effect of the companion planet, we also ran one control simulation under the same conditions outlined above, but with no planet. This integration shows the star's rotational evolution in the absence of planet-induced tidal evolution, yielding a rotation rate of 6.5 days after 500 Myr. However, for each individual simulation, the behavior of the planet affects the observed rotation rates. Two examples of this effect can be seen in Figure \ref{fig:Ng1}. In both panels, the control simulation is given as a dashed line, and the red line shows the evolution of the stellar rotation as computed in the presence of the planet. A second line (right y-axis, tan line) shows the semi-major axis evolution of the planet. {These examples were chosen as representative of each of their behaviors, as described in the caption on the figure.}
Two two panels show ways in which the planet can increase or decrease the stellar rotation rate. In the upper panel, the planet's orbit expands as it extracted angular momentum from the star and the stellar rotation period slows compared to the no-planet integration. In the bottom panel, the planet experiences quick tidal in-spiral immediately after the disk dissipates and it is subsequently engulfed by the star. This results in it depositing its angular momentum onto the star and increasing the stellar rotation rate as compared to the no-planet case. 

\begin{figure*}
    \centering
    \includegraphics[width=0.95\linewidth]{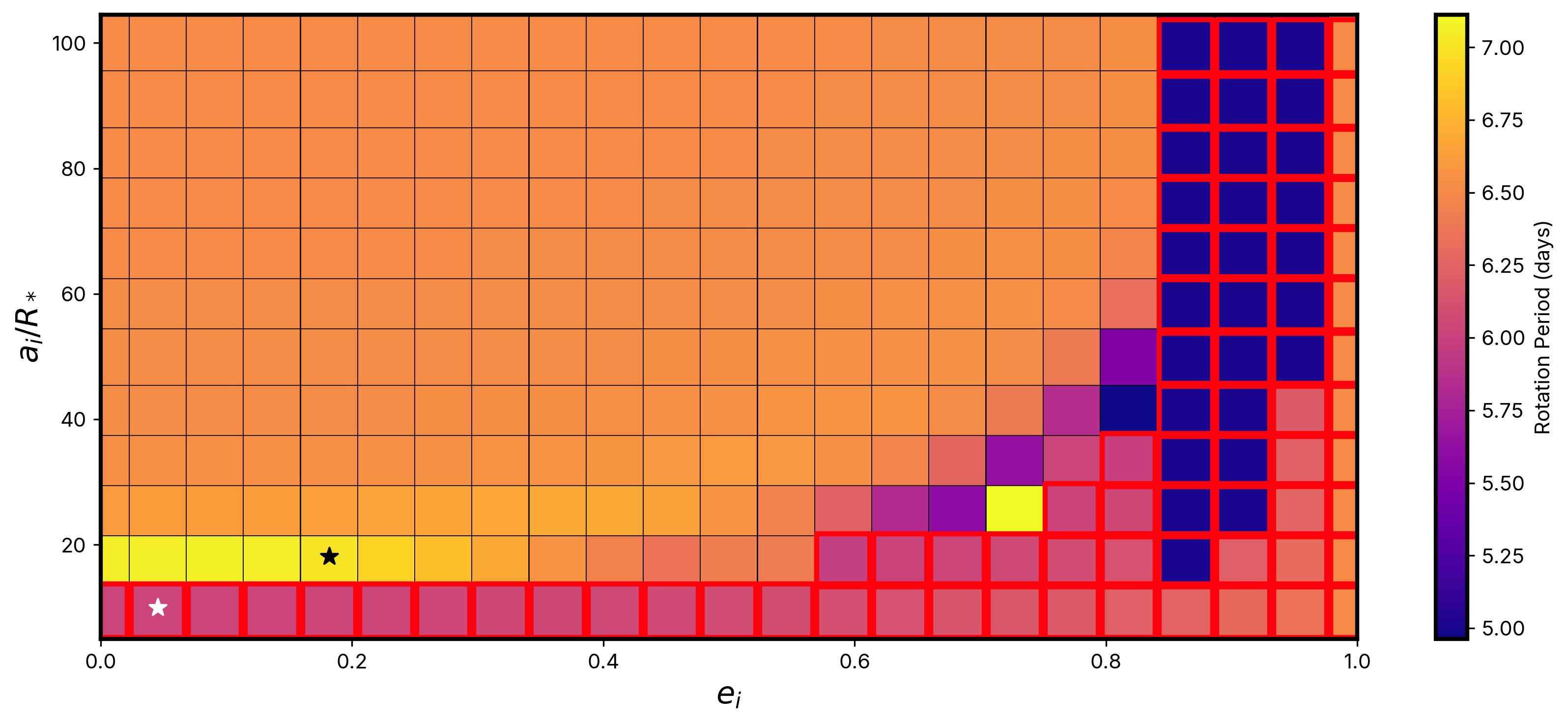}
    \caption{A grid showing the rotational period of the star computed by \poet\ after 500 Myr given a range of initial semi-major axis (measured in units of $a/R_*$) and eccentricity conditions for the planet. Thicker red borders around some cells indicate that those combinations of planetary initial conditions led to the planet being engulfed. Those initial conditions are not consistent with observations since no planet would be discovered at current day. For simulations where the planet survives, the stellar rotation period ranges between 5 and 7 days, as compared to the 6.5 day rotation rate that the simulation would show if no planet were present. {The two stars represent the example cases shown in Figure \ref{fig:Ng1}, with the black star representing the $a/R_* = 18.18$ and $e = 0.18$ `surviving' case (top panel of Figure \ref{fig:Ng1}) and the white star representing the $a/R_* = 10.0$ and $e = 0.0$ `engulfment' case (bottom panel of Figure \ref{fig:Ng1}).}}
    \label{fig:spinupgrid}
\end{figure*}

The results of the full suite of 300 simulations are shown in Figure \ref{fig:spinupgrid}. 
Our simulated planets followed a variety of tracks in their orbital evolution, including being engulfed by their host. 
We note that due to the coarseness of our sampled grid, none of our simulations had planets that ended the simulation at a semi-major axis matching that of that of \pname\ ($a/R_* \approx 3.8)$. As such, our simulations do not recreate the exactly geometry of \name, but rather compute the range of alterations to stellar rotational period that could be possible due to interactions with a planet with \pname's parameters. For simulated systems where the planet avoided being engulfed by the star, the stellar rotation period ranged between a minimum of 5 - 5.75 days (for high-$e$, low-$a$ initial conditions) and a maximum of 7 days (for low-$e$, low-$a$ initial conditions). Low-$e$ and high-$a$ planet parameters resulted in no alteration to the stellar rotation period as compared to the 6.5 days seen in the no-planet case. 

Our initial, limited set of simulations does not encompass the entire range of possible initial parameters for this system. We have used only one value of initial spin angular momentum $L_{*}$ for the star and simplified the problem by assuming constant tidal quality factors for the star and the planet. However, for this particular set of simulations, it is evident that the effect of the planet-induced tidal evolution over the first 500 Myr of evolution is to change the inferred stellar rotation by (at maximum) a couple of days. Depending on the exact initial parameters, this change can either spin up the star, resulting in a faster rotation period, or, for a small range of initial parameters, it can slow the stellar rotation. Given the observed orbit of this planet, it is unlikely that it has slowed the stellar rotation int the past, as it would have had to in the past have resided on an even closer orbit as compared to its current orbit. It is more likely that the effect of the planet has been to increase the star's rotation compared to what it would have been if no planet were present.

\subsection{Dynamical Pathways for \pname}
While our simulations in the previous subsection demonstrated the range of possible effects of tidal evolution due to a massive planet like \pname, our simulations were not sufficiently focused to answer the question: what is the future tidal evolution of \pname\ expected to look like?
In order to probe the potential dynamical future of \pname, we ran seven simulations initialized to the current state of \name, including all the best-fit parameters reported in Table \ref{tab:params} ($a/R_* = 3.8$, $e=0$, $R_p = 1.247 R_{J}$, $M_p = 2.0 M_{J}$, $M_* = 1.072 M_{\odot}$, and a stellar rotation rate of 6.5 days at 500 Myr, as assumed in the previous section). The simulation was initialized at 500 Myr with those parameters, and the only value changed between simulations was the stellar tidal quality factor $Q_*$. $Q_*$ is a parameter expected to vary significantly across physical and orbital \citep{Patel2023} regimes. It cannot be measured directly, and must instead be inferred from dynamical constraints, such as planet survival \citep{Hamer2020}.

\begin{figure}
    \centering
    \includegraphics[width=0.45\textwidth]{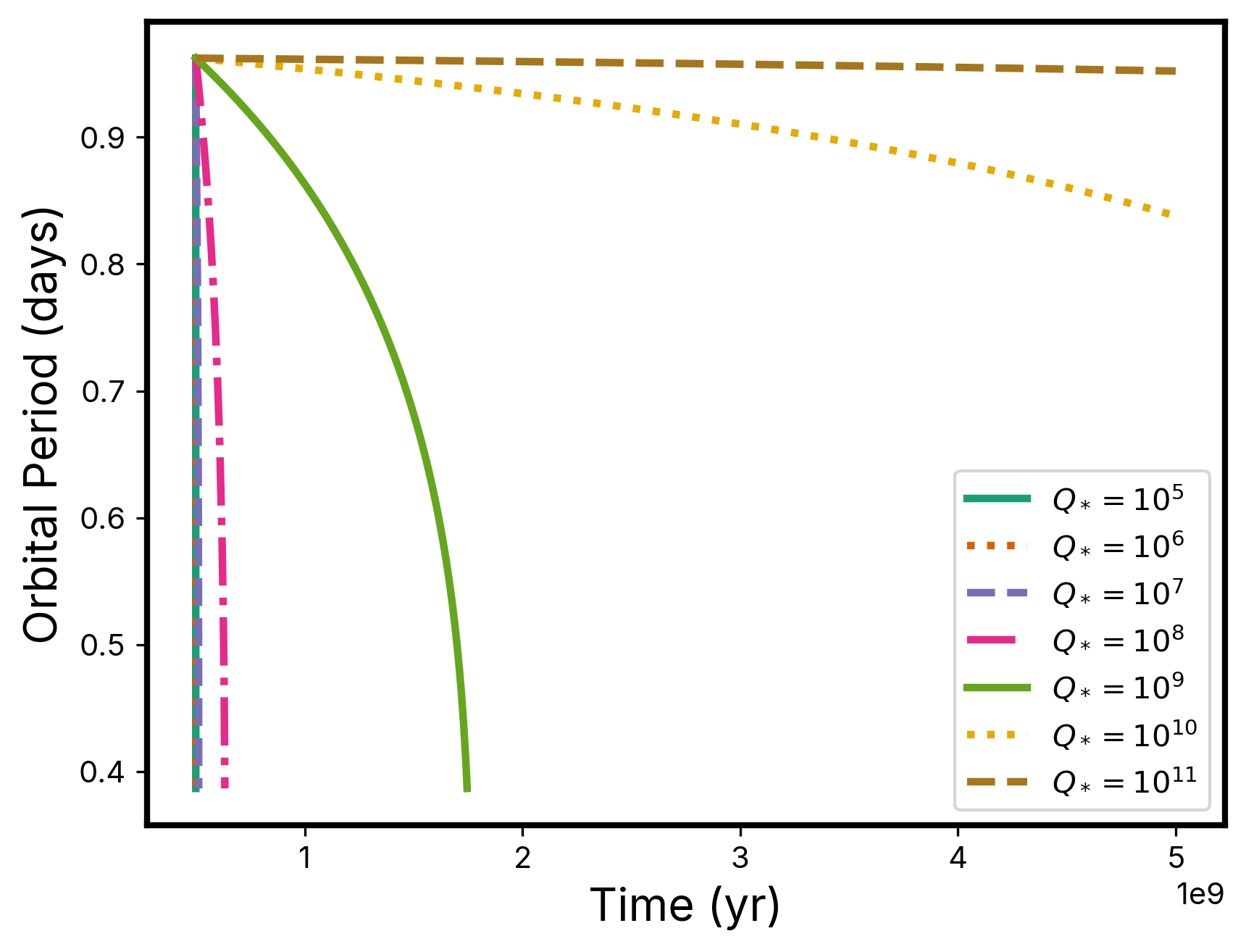}
    \caption{Orbital evolution of TOI-1937Ab's orbital period over time for various values of $Q_*$, illustrating the impact of tidal dissipation efficiency on future orbital evolution of the planet. Higher $Q_*$ values correspond to slower orbital evolution, while lower $Q_*$ values correspond to more rapid decay. $Q_*< 10^8$ correspond to engulfment of TOI-1937Ab in the near future.}\label{fig:Qorbit}
\end{figure}

We tested values of $Q_*$ logarithmically spaced between $10^5$ and $10^{11}$ and ran integrations for 4.5 Gyr (from a starting time of 500 Myr to a end time of 5 Gyr). The results of these simulations are plotted in Figure \ref{fig:Qorbit}. All seven integrations show that \pname\ is currently in a state of tidal decay. However, the different values of $Q_*$ indicate different likely futures for this planet: for $Q_*\le10^7$, the planet is expected to be engulfed by the star within the next 500 Myr; for $10^8 \approx Q_*$, the planet's orbit will decay and it will be engulfed in slightly over 1 Gyr from now; for $10^9 \le Q_*$, the planet will not be engulfed within 5 Gyr. 

\begin{figure}
    \centering
    \includegraphics[width= 0.45\textwidth]{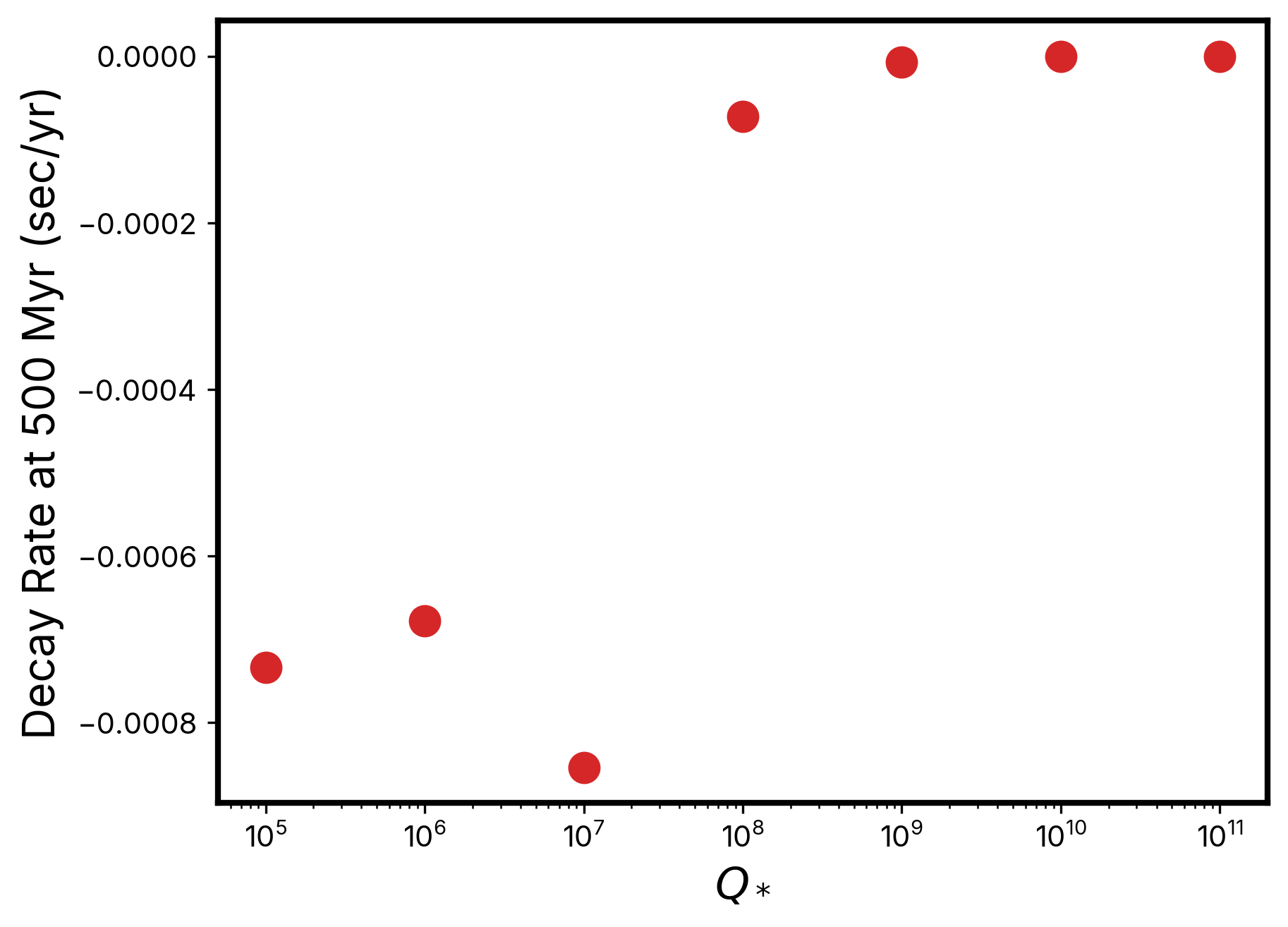}
    \caption{The computed decay rate of the planet's orbital period ($\dot{P}$) at 500 Myr, as it depends on $Q_*$, computed from our \poet\ simulations for each tested value of $Q_*$. The largest computed decay rates occur for the smallest values of $Q_*$. With a sufficiently stringent limit on the observed value of $\dot{P}$ from the TTVs, we would be able to place a constraint on the stellar $Q_*$; however, the upper limit on the decay rate computed from the TTVs is higher than all theoretically expected decay rates shown here.}
    \label{fig:Qdecay}
\end{figure}

Next, we computed the instantaneous decay rate of the orbital period, with a goal of comparing how these theoretically-derived decay rates compared to the upper limit on the decay rate computed in Section \ref{sec:two}. {The results of these calculations are plotted in Figure \ref{fig:Qdecay}.}
However, none of our simulated planets had decay rates that were greater than our derived upper limit on the decay rate for \pname. For this reason, a concrete constraint on the $Q_*$ of \name\ cannot be made with currently available data. 

\section{Discussion} \label{sec:discussion}

{\subsection{Implications of Tidal Evolution Simulations and TTV Analysis}}
Our simulations provide partial evidence suggesting that this planet attained its observed orbit recently, given that, unless the host star's $Q_*$ is anomalously high, it would have been engulfed by its host star relatively quickly. However, this cannot be taken as evidence on the host star's age, as it is possible that the dynamical interaction that placed \pname\ on its high-periastron orbit (which allowed for the start of the tidal evolution process) happened recently, even if the star is older. 
Additionally, our \poet\ simulations to asses the rotational period alterations caused by the planet's tidal evolution suggest that the change in rotational period for the host star is marginal (or the order of 1-2 days, which is commensurate with the $3\sigma$ errors on the rotation rate) unless the planet is engulfed by its host star. 

From our TTV analysis, we find no evidence of a non-linear ephemeris, and find \edittwo{a $3\sigma$ limit on the orbital period decay rate of \ttvestimate.} Additionally, our TTV analysis shows no evidence of additional planets near \pname. {Since it is expected} that this planet migrated via tidal migration—a process that excites higher orbital eccentricity and could lead this planet to cross the orbits of any other companion planets \citep{Dawson2018}—it is not surprising that we find no evidence of nearby companions. The majority of hot Jupiters are observed not to have close planetary companions. However, the absence of detected companions does not necessarily mean they are not present. Due to the large measured mass of this planet, any nearby planetary companions would have to be relatively large to cause observable TTVs.

We also considered the tidal evolution of \pname. \pname\ is a hot Jupiter in an ultra-short period orbit, one of only roughly eleven similar planets in the exoplanet census. Most observed USP planets are less than 2 $R_{\odot}$ \citep{Winn2018}, and population-wide evidence suggests that hot Jupiters are subject to significant tidally-driven orbital decay on the main sequence \citep{Hamer2019}. While the age of the systems remains amibiguous, if \name\ were a young ($<1$ Gyr) host star, it would place a useful constraint on the earlier arrival of hot Jupiters to their orbits. 
We also note that our \poet\ simulations, presented in Section \ref{sec:tides}, demonstrate that it is likely that \pname\ is in actively decaying orbit, with its final time of engulfment set by the stellar $Q_{*}$. For smaller ($\approx 2.6 M_{Earth}$) USP planets, \citet{Hamer2020} finds that stars need to have $Q_{*} > 10^7$ for the planets to be stable against tidal inspiral as a population. For \pname, a $Q_* \approx 10^{10-11}$, a value far above the median $Q_*$ for smaller USP planets found in \citet{Hamer2020}, {would protect} the planet against engulfment on the main sequence.

\subsection{{TOI-1937 B and Kozai-Lidov Interactions}}
{The presence of the stellar companion TOI-1937 B at a projected separation of approximately 1030 AU \citep{Yee2023} raises the possibility of dynamical interactions that could influence the orbit of \pname. 
One potentially relevant interaction is the Kozai-Lidov mechanism \citep{Lidov1962, Kozai1962}, which can induce large oscillations in a planet's orbital eccentricity and inclination, potentially setting the stage its subsequent tidal migration into its final observed short-period orbit. While the full orbital parameters of the stellar companion (TOI-1937B) - including its mass, eccentricity, and semi-major axis - are not well constrained, we performed an analysis to evaluate whether the Kozai-Lidov mechanism could have contributed to the formation of TOI-1937Ab’s current orbit.}

{As the orbital and physical parameters of companion TOI-1937 B are not known, for this analysis we assumed a companion mass of 0.3 $M_{\odot}$\edittwo{, approximated based on the brightness difference between the primary and secondary and consistent with the results derived in \citet{Christian2024},} and assumed that its measured projected separation of 1030 AU is its semi-major axis. The Kozai-Lidov timescale can be computed by (see Equation 42 of \citealt{Antognini2015}; see also \citealt{Holman1997, Christian2022}):}
\begin{equation}
    \tau \simeq \frac{8}{15\pi} \left( 1 + \frac{m_1}{m_3} \right) \left( \frac{P_{\mathrm{out}}^2}{P_{\mathrm{in}}} \right) (1 - e_2^2)^{3/2} ,
    \label{eq:kozai}
\end{equation}
{where $m$ denotes masses, $P$ orbital periods, $e$ orbital eccentricity, and subscripts $1,2,3$ denote the primary star, planet, and stellar companion, respectively.
We computed the Kozai-Lidov timescale over a range of plausible initial orbital periods and eccentricities for TOI-1937Ab, prior to any scattering or migration event. The results are presented in Figure \ref{fig:kozai}, which shows Kozai-Lidov timescales for different initial conditions of the planet’s orbit. To contextualize these timescales, we overlaid contour lines corresponding to the estimated age of the system.}

{Only orbital configurations to the right of these lines have sufficiently short Kozai-Lidov timescales to operate within the system's lifetime. \edittwo{In Figure \ref{fig:kozai}, we over-plot grey points corresponding to the cold Jupiters ($M_p > 0.5 M_{jup}$) discovered via radial velocity observations.\footnote{Data obtained from the IPAC Exoplanet Archive, 2/17/2025.} These points populate the parameter space where the Kozai-Lidov timescale would be short enough to be expected to operate for all feasible system ages. As a result, we consider it} dynamically feasible that \edittwo{if \pname\ had started as a typical cold Jupiter from the RV sample \citep[i.e.,][]{Bryan2016, Wittenmyer2020, Rosenthal2021}, TOI-1937B could have played a role in setting TOI-1937Ab on an initial} eccentric orbit. 

However, we note that for the mechanism to result in the observed hot Jupiter orbit, there are additional conditions beyond the Kozai timescale being short enough to operate within the stellar lifetime: the planet's periastron distance would need to become sufficiently small to allow for significant tidal dissipation and circularization within the system's lifetime. Additionally, a Rossiter-McLaughlin measurement by \citet{Yee2023} found that the orbit of the hot Jupiter \pname\ is roughly aligned with the spin axis of the host star \name. This would suggest that \pname\ either migrated via coplanar high-eccentricity migration \citep{Petrovich2015} or had its obliquity damped post-migration \citep{Lai2012, Spalding2022}. While these additional factors require further investigation, our analysis supports the plausibility of Kozai-Lidov cycles as a contributing mechanism in the evolutionary history of TOI-1937Ab.

\begin{figure}[h!]
    \centering
    \includegraphics[width= 0.45\textwidth]{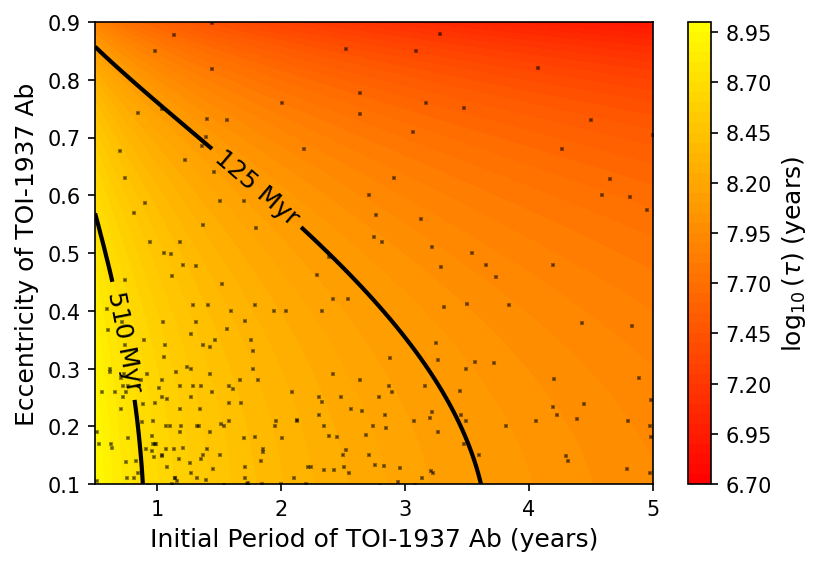}
    \caption{The predicted Kozai-Lidov timescale (Equation \ref{eq:kozai}) for a range of initial planet parameters. Any set of parameters above a line has a Kozai-Lidov timescale short enough to bring a Jupiter with those initial parameters into a high-eccentricity orbit by the noted time. A large range of initial planet parameters result in Kozai-Lidov timescales short enough to operate by even the shortest possible age of \name\ (125 Myr). \edittwo{We overplot as grey squares the cold Jupiter population obtained from the IPAC Exoplanet Database.} \edittwo{We note that, were the companion star to have a higher mass, the contours would shift to the left. Were the companion star to have a lower mass, the contours would shift to the right. In either case, the Kozai-Lidov timescale is not prohibitively long for this type of migration to have occurred.} }
    \label{fig:kozai}
\end{figure}

{\subsection{The Possible Ages of \name}}
{The age estimates computed in this work for \name\ are presented in Table \ref{tab:ages}. Different pieces of dynamical and physical evidence give age estimates. The gyrochronological age estimate is around 510 Myr, the Li abundance gives a $1 \sigma$ lower limit of roughly 4.5 Gyr,  and were \name\ a member of the NCG 2516 cluster it would be roughly 150 Myr old. }
The non-detection of lithium, described in \citet{Yee2023}, remains the most confounding piece in age dating this star. While this star is likely not part of the NGC 2516 open cluster, as previously suggested in \citet{Kounkel2019}, much of the evidence, with the exception of lithium abundance, points to this star being younger than 1\,Gyr. In \citet{Sevilla2022}, it is suggested that a planetary engulfment could decrease surface lithium abundance in more massive stars ($1.3-1.4$\Msun) as a result of an enhancement of internal mixing and diffusion processes, however, TOI-1937A is not in this mass regime. Thus, we believe that the lithium-derived age estimate is inconclusive and further investigation and observations are required to confidently assign an age for this system.
\linebreak

\subsection{Avenues for Future Work}
\subsubsection{TTVs}
The TTV precision we obtained with the presently available TESS data and follow-up ground-based data allowed us to placed a upper limit on the decay rate of the orbital period of \pname. However, the data was not of a sufficient precision or observational baseline to find (or exclude) a concrete detection of orbital decay.  
In order to improve our constraint on this decay rate and provide a constraint on $Q_*$, more transit observations are necessary. 
Currently, the available observations span around $\sim$1600 transit epochs of \pname, but the errors on transit timing for the earliest data available (the TESS FFI data) are on the order of a few minutes. 
The TTV analysis can be improved in two ways: additional data at a longer observational baseline and improved timing precision on future events. 

This will be possible in the future with additional data taken with TESS or with ground-based resources. Even in the relatively short history of exoplanet transit observations, it has been demonstrated that significantly improved constraints on decay rate are possible once the observation baseline reaches a decade or more. For example, early studies of WASP-19b \citep{Mancini2013} indicated that it might have a decay rate of -10 sec/yr or more \citep{Espinoza2019}, but years of additional data brought this estimate down to an upper limit on decay rate of -2.294 ms/yr \citep{Petrucci2020}. \pname\ is an excellent target for ground-based follow-up observations because, as shown in Figure \ref{fig:TTVs}, its large radius and high SNR result in surprisingly good transit timing precision for ground-based data. In the future, with a large enough baseline between the existing transit data and additional observed transit events, we expect that a significantly improved constraint on the orbital period decay rate of \pname\ will be possible. 

\subsubsection{Numerical Modeling of the Tidal Evolution}
In Section \ref{sec:tides} of this paper, we present two sets of simulations using the \poet\ code that explore the potential tidal evolution of this planet, first from formation to present day and then predicting the future evolution of the system based on the current observed state of the system. The first set of simulations, aimed at determining the range of stellar rotation periods possible in the presence of planet-induced tidal evolution, do not perfectly reproduce the present-day orbit of this planet ($a/R_* = 3.8$, $e = 0$) within the 500 million year integration time. 
One interesting next step would be to directly constrain the orbital parameters of the planet that are consistent with producing the currently observed parameters. This would provide a range of possible semi-major axis and eccentricity values that the planet could have had prior to the scattering event that increased its eccentricity and began the tidal migration process in the past.
Additionally, future numerical explorations would benefit from testing the other possible ages for \name, as given in Table \ref{tab:ages}. 
The simulations presented in Section \ref{sec:tides} assume an age of 500 million years, consistent with estimates from gyrochronology. However, it is also possible that the star is significantly older \citep[as suggested from the Lithium constraint of this work and the EXOFASTv2 estimate from][]{Yee2023} and has been spun up due to interactions with the planet, in which case the tidal evolution would unfold over a longer timescale than we are currently modeling. Finally, running a finer grid than that presented in Figure \ref{fig:spinupgrid} would allow for a determination of the likely initial semi-major axis and eccentricity values of this planet, even within the framework of our current simulation parameters.

\subsubsection{Investigations Into \pname's Radius}
This planet is a hot Jupiter with a very high level of stellar irradiation. Planets of this type are often seen to have inflated radii \citep{Laughlin2011, Demory2011} compared to what would be expected for a colder planet for the same mass.
\pname\ is an ultra-hot Jupiter, subject to tidal forcing \citep{Leconte2010} and with a high enough level of irradiation \citep[with a flux $F \approx 4.4 \times10^9$ erg s$^{-1}$ cm$^{-2}$;][]{Yee2023} that it should also be subject to ohmic dissipation \citep{Batygin2010, Thorngren2018}, indicating that its radius would be expected to be inflated. 
However, when compared to the population of irradiated hot Jupiters as a whole \citep[see for example Figure 2 of][]{Thorngren2016}, \pname\ resides at the lower edge of the radius distribution for planets of its surface irradiation flux. 
Even if \pname\ arrived on its observed orbit recently, the inflation process is fast enough that it should be inflated \citep{Thorngren2021}. 
This requires further investigation, including observations or numerical circulation models \citep[i.e.,][]{Komacek2022} to better determine whether \pname\ is simply the lower edge of the statistical distribution or if some additional effect explains its non-inflated radius. 

\section{Conclusion} \label{sec:conclusion}
Our analysis provides new insights into the dynamical history and potential future evolution of the \name\ system. \pname\ represents a valuable test case for theories of planetary migration, tidal dissipation, and stellar-planet interactions due to its high mass and ultra-short-period orbit.  

Using a transit timing variation analysis and tidal evolution simulations with \poet, we find that \pname, an ultra-short-period hot Jupiter, is {theoretically} likely undergoing orbital decay driven by tidal interactions. While no statistically significant orbital decay was detected {observationally}, \edittwo{we place an upper limit on the decay rate of $\dot{P} < $ \ttvestimate.} Our simulations demonstrate that the planet's future evolution will depend on the stellar tidal quality factor $Q_*$, with typical stellar values leading to the planet's engulfment within 500 Myr of the present day.

Our analysis raises questions about the system's age, with discrepancies between gyrochronological and lithium abundance estimates. While the lack of lithium supports an older stellar age, tidal spin-up caused by \pname\ complicates gyrochronological estimates. 
Based on the observed orbital dynamics and stellar lithium abundance, we believe that it is unlikely that \name\ is a true member of the NGC 2516 open cluster as previously proposed. 
Further observations are needed to resolve these uncertainties.




\vspace{4mm}
We would like to thank Samuel Yee for sharing relevant data and code.
We would also like to thank Joshua Schussler for help getting \poet\ running and Ellen Price for debugging help. 
{We would also like to thank the referee for their helpful comments.}
We thank Coco Zhang for useful conversations. 
ARJ would like to thank the generous support of the Wisconsin Space Grant
Consortium under NASA Award No. 80NSSC20M0123 and the University of Wisconsin Physics Department through the Hubert Mack Thaxton Fellowship.
\edittwo{This research has made use of the NASA Exoplanet Archive, which is operated by the California Institute of Technology, under contract with the National Aeronautics and Space Administration under the Exoplanet Exploration Program.}

%

\vspace{5mm}
\facilities{Mikulski Archive for Space Telescopes \citep{MAST}, TESS, Exoplanet Archive}


\software{
\texttt{BAFFLES} \citep{baffles}, 
\texttt{batman} \citep{Kreidberg2015}, 
\texttt{corner.py} \citep{cornerpy},
\texttt{emcee} \citep{Foreman-Mackey2013}
\texttt{gyrointerp} \citep{Bouma2023}, 
\texttt{matplotlib} \citep{Hunter:2007},
\texttt{pandas} \citep{mckinney-proc-scipy-2010, the_pandas_development_team_2024_13819579},
\texttt{POET} \citep{Penev2014software}, 
          }

\bibliography{PASPsample631}{}
\bibliographystyle{aasjournal}



\end{document}